# Acceleration of 60 MeV proton beams in the commissioning experiment of SULF-10 PW laser

A. X. Li[1,2,†], C. Y. Qin[1,3,†], H. Zhang[1,4, a)], S. Li[1], L. L. Fan[1,3], Q. S. Wang[1,5],

T. J. Xu[1], N. W. Wang[1], L. H. Yu[1], Y. Xu[1], Y. Q. Liu[1], C. Wang[1], X. L. Wang[1],

Z. X. Zhang[1], X. Y. Liu[1], P. L. Bai[1], Z. B. Gan[1], X. B. Zhang[1], X. B. Wang[1], C. Fan[1],

Y. J. Sun[1], Y. H. Tang[1], B. Yao[1], X. Y. Liang[1,4], Y. X. Leng[1,4], B. F. Shen[1,6, b)],

L. L. Ji[1,4, c)], and R. X. Li[1,2,4, d)]

[1]*State Key Laboratory of High Field Laser Physics, Shanghai Institute of Optics and Fine Mechanics, Chinese Academy of Sciences, Shanghai 201800, China*

[2]*ShanghaiTech University, Shanghai 201210, China*

[3]*Center of Materials Science and Optoelectronics Engineering, University of Chinese Academy of Sciences, Beijing 100049, China*

[4]*CAS Center for Excellence in Ultra-intense Laser Science, Shanghai 201800, China*

[5]*College of Science, University of Shanghai for Science and Technology, Shanghai 200093, China*

[6]*Department of Physics, Shanghai Normal University, Shanghai 200234, China*

We report the experimental results of the commissioning phase in the 10 PW laser beamline of Shanghai Superintense Ultrafast Laser Facility (SULF). The peak power reaches 2.4 PW on target without the last amplifying during the experiment. The laser energy of 72±9 J is directed to a focal spot of ~6 μm diameter (FWHM) in 30 fs pulse duration, yielding a focused peak intensity around $2.0\times10^{21}$ W/cm$^2$. First laser-proton acceleration experiment is performed using plain copper and plastic targets. High-energy proton beams with maximum cut-off energy up to 62.5 MeV are achieved using copper foils at the optimum target thickness of 4 μm via target normal sheath acceleration (TNSA). For plastic targets of tens of nanometers thick, the proton cut-off energy is approximately 20 MeV, showing ring-like or filamented density distributions. These experimental results reflect the capabilities of the SULF-10 PW beamline, e.g., both ultrahigh intensity and relatively good beam contrast. Further optimization for these key parameters is underway, where peak laser intensities of $10^{22}$-$10^{23}$ W/cm$^2$ are anticipated to support various experiments on extreme field physics.

———————————————

† These authors contributed equally to this work.

a) Email: zhanghui1989@siom.ac.cn

b) E-mail: bfshen@mail.shcnc.ac.cn

c) E-mail: jill@siom.ac.cn

d) E-mail: ruxinli@mail.siom.ac.cn

# 1. Introduction

Chirped-pulse amplification (CPA) technology significantly advanced the development of ultrashort ultraintense laser in the past 37 years [1-3]. Today nearly one hundred 100 TW systems are operating with about 20 systems at the PW level existing or under construction [3], pushing laser intensities to go beyond the relativistic threshold (about $10^{18}$ W/cm$^2$ for laser wavelength of ~μm). Unprecedented extreme physical conditions can be created in the lab [4-6], which strongly motivate the studies of laser-driven particle acceleration[4,7,8], x/gamma ray radiation [9-11], laboratory astrophysics [6,12,13], laser-driven nuclear physics [14] etc. On the other hand, the rising interest in strong-field quantum electrodynamics calls for lasers with even higher intensities ($10^{22}$~$10^{23}$ W/cm$^2$). Such quest has been supported by several projects aiming to reach 10 PW-level outputs, such as ELI [15], Vulcan-10 PW [16], Apollon-10 PW [17], and SULF-10 PW [18]. The first 100 PW-level laser facility under construction is the Station of Extreme Light Science (SEL) [19], while several others are also under consideration (ELI-200 PW, XCELS, Nexawatt, Gekko EXA, Rochester etc).

The Shanghai Superintense Ultrafast Laser Facility (SULF) is the first 10 PW-class laser facility in China, which was proposed and constructed by the Shanghai Institute of Optics and Fine Mechanics (SIOM) in July 2016. Figure 1 shows the layout of SULF laser facility [20]. The SULF laser employs typical CPA Ti: sapphire scheme and contains two high-intensity laser beamlines, SULF-10 PW operating at repetition rate of one shot per three minutes [21] and SULF-1 PW operating at repetition rate of 0.1 Hz [22]. In 2018, the SULF-10 PW beamline has realized output peak power up to 10.3 PW (after compressed) with 339 J output pulse energy (compressor transmission efficiency of 64%) compressed to 21 fs pulse duration [18]. This peak power was further increased to 12.9 PW in 2019 [21]. The physical experimental areas in SULF include three research platforms of dynamics of materials under extreme conditions (DMEC), ultrafast sub-atomic physics (USAP) and big molecule dynamics and extreme-fast chemistry (MODEC).

The commissioning experiment of SULF-10 PW beamline was carried out on the USAP platform, focusing on laser-proton acceleration using the plain Cu and plastic targets. The peak power reaches 2.4 PW without the last amplifying section, corresponding to laser energy of 72±9 J, focal spot size of ~6 μm diameter (full-width-half-maximum, FWHM) and 30 fs pulse duration. These together yield a focused peak intensity around $2.0 \times 10^{21}$ W/cm$^2$. We obtained proton beam with cut-off energy up to 62.5 MeV using Cu target at the optimum target thickness of 4 μm. For much thinner targets (tens of nanometers), the proton cut-off energy declines to 20 MeV and ring-like or filamented structures appear in the density distribution. The obtained results from laser-foil interaction directly illustrate the current capabilities of the SULF-10 PW beamline.

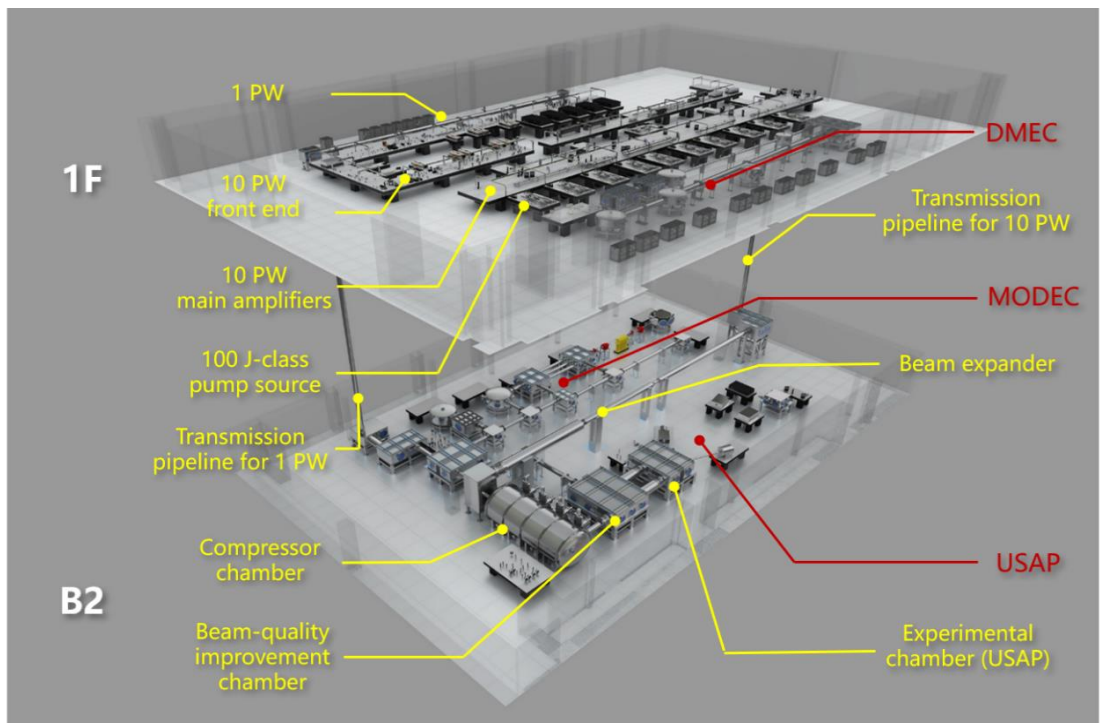

Fig. 1. The layout of SULF laser facility [20].

## 2. The current status of experimental area and SULF-10 PW beamline

**2.1 Experimental area in USAP**

As seen in Fig. 1, the SULF-10 PW laser beam go through 7 multi-pass amplifiers at the first floor of the SULF building, and then is transmitted to B2 floor through transmission pipeline. Following that the amplified beam is further expanded and image relayed into the compressor cavity. Behind the compressor chamber are the beam-quality improvement chamber and experimental chamber (USAP). Figure 2 shows a photograph of the experimental area in USAP located on the B2 floor. The beam-quality improvement chamber is specially designed to place deformable mirror (DM) and plasma mirrors (PM) to further improve the beam quality and contrast of laser. The experimental chamber for laser-matter interaction comprises two vacuum cavities, the larger one for short-focal-length experiments like laser-ion acceleration while the smaller one for experiments requiring long focal length such as laser wakefield acceleration of electron. The internal size of both the beam-quality improvement and the short-focal-length chambers is 4.5 m×3.5 m×2.0 m and it takes about one hour for the vacuum system to pump it from standard atmospheric pressure down to $10^{-4}$-$10^{-3}$ Pa. An important function of the USAP platform is that it allows users to simultaneously employ both the SULF-10 PW and SULF-1 PW beamlines, for either pump-probe or laser-electron scattering experiments. In the commissioning experiment on the SULF-10 PW beamline, only the off-axis parabola (OAP) of short focal length and the DM in the beam-quality improvement chamber were employed without introducing the PMs in the laser path.

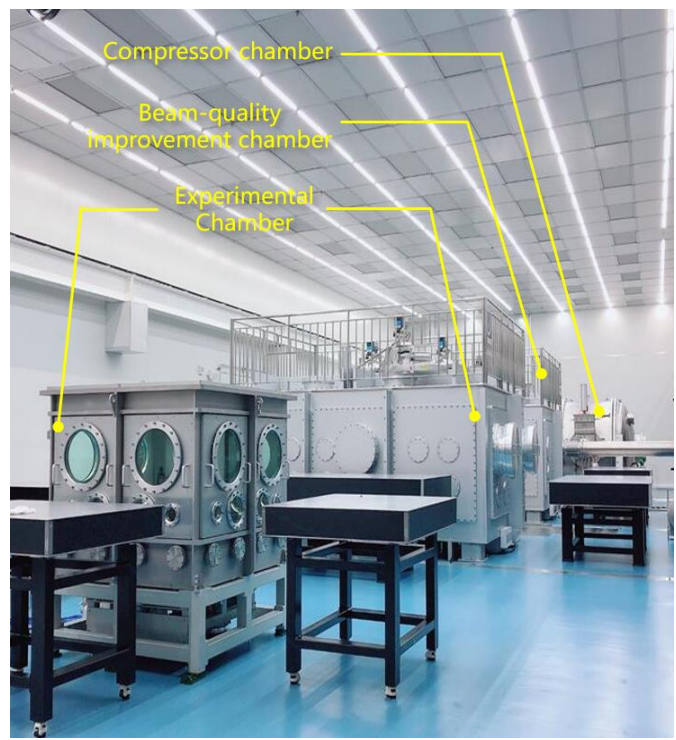

Fig. 2. The experimental area in USAP.

## 2.2 Laser parameters of the SULF-10 PW beamline

In the commissioning stage, the output energy of the SULF-10 PW laser before the compressor is measured to be 110±13 J with 6 multi-pass amplifiers (the last one was switched off). The energy transmission efficiency from the compressor to target is 66%, resulting in the on-target energy of 72±9 J. The near-field profile of the final output laser is elliptic, 470 mm×430 mm large along horizontal and vertical directions. The measured modulation of the near-field beam is about 1.8, mainly due to the modulation of the pump laser beam from the main amplifier.

Decreasing the size of the focal spot is an efficient method to increase the laser intensity. However, the large-aperture optic elements implemented in the SULF-10 PW system inevitably increases the wavefront aberrations. Here, double DMs with different actuator densities are cascaded to optimize the wavefront aberrations and hence the focal intensity [23]. The first adaptive-optics (AO) correction system is placed at the output of the 6th amplifier, and the DM has a diameter of 130 mm with 64 mechanical actuators. A second AO correction loop is installed using a larger DM with 520 mm diameter and 121 mechanical actuators. It is placed at the output of compressor (in the beam-quality improvement chamber). The wavefront sampling laser is exported out of the experimental chamber following the light path built after an OAP. The focal spot is amplified by 10 times and then online monitored by a low-noise CCD. An OAP with 2000 mm focal length is used for laser-proton acceleration, corresponding to effective $f$-number of 4.4. Figure 3(a) shows the typical focal laser intensity distribution after the correction. The FWHM size of focal spot is 6.28 μm×5.92 μm, containing 24% of the total laser energy.

The measured spectral width of the output pulse is ±40 nm (FWHM) at 800 nm central wavelength. The laser beam is compressed by a four-grating compressor and the pulse duration is measured using a Fastlite Wizzler instrument. Figure 3(b) shows that the typical pulse duration is about 30 fs (FWHM), resulting in an output peak power of 2.4 PW. These data indicate that the focal peak intensity reaches $2.0\times10^{21}$ W/cm$^2$.

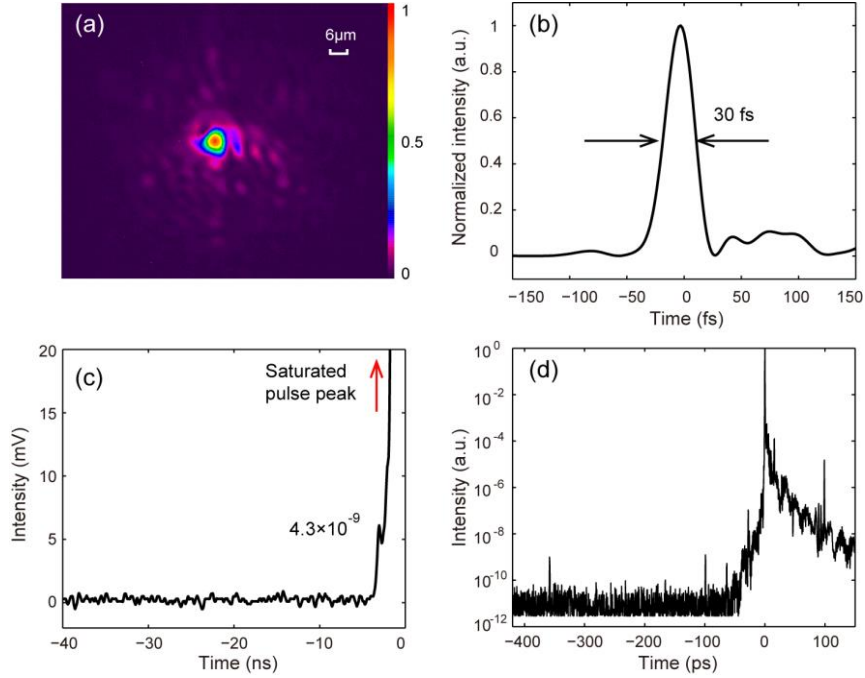

Fig. 3. Laser parameters of SULF-10 PW beamline for the commissioning experiment. (a) The typical focal spot of the laser after the correction of double DM system, which is measured using a low-noise CCD by the light path built after an *f*/4.4 OAP. (b) The typical pulse duration of the compressed pulse measured by a Fastlite Wizzler instrument. The temporal contrast of (c) nanosecond scale measured by a photodiode with a stack of neutral attenuator and (d) picosecond scale measured by a third-order cross-correlator. The red arrow represents the saturated peak of the laser pulse.

A key parameter for laser-solid interaction at relativistic intensities is the temporal contrast of the laser pulse. Pre-pulses of intensity above $10^{10}$ W/cm$^2$ [24] would ionize the target, introducing low-density pre-plasmas in front of the target which could enhance proton acceleration[25,26]. However, if the pre-plasma induced by pre-pulse driven shock appears at target rear, proton acceleration will be restricted[27]. In order to improve the temporal contrast of SULF-10 PW beamline, a combination of cross-polarized wave generation (XPWG) and femtosecond optical parametric amplification (OPA) technique is implemented at the front end [28]. Along with further optimization [29], the contrast ratio at 50 ps before the main pulse is measured to be around $1.7 \times 10^{-9}$. Meanwhile, two Pockels cells are installed after the first amplifier to increase nanosecond temporal contrast. Here, the contrast evolution at nanosecond scale was measured by a combination of oscilloscope and photodiode at the output of the compressor as shown in Fig. 3(c). It can be found that the amplified spontaneous emission (ASE) noise level is better than $10^{-9}$ (limited by the photodiode). An intense pre-pulse is also seen at -3 ns with the contrast of $\sim 4.3 \times 10^{-9}$ for reasons that are still under investigation. The temporal contrast at the picosecond scale was measured by a commercial third-order cross-correlator (Amplitude, Sequoia). The contrast curve within -420 ps before the main pulse is illustrated in Fig. 3(d), showing a pedestal around $10^{-11}$, which starts rising from -50 ps. In addition, three pre-pulses appear at -360 ps, -100 ps and -60 ps with the contrast ratio of $\sim 10^{-9}$ due to multiple reflections of the optical components in the amplifiers (The detailed analysis of pre-pulses will be introduced in another paper). Considering the laser intensity of $10^{21}$ W/cm$^2$, these pre-pulses reach intensities of $10^{12}$ W/cm$^2$, sufficiently strong to trigger material ionization, and could induce pre-plasmas on target rear. Plasma mirrors are to be installed in the near future to improve the performance on laser-foil interactions.

## 3. Experiment results

### 3.1 Experimental setup

The sketch of the experimental setup is shown in Fig. 4. The *p*-polarized laser pulse is focused by the *f*/4.4 OAP mirror onto the target at an incident angle of 15°. In this run, the target table accommodates 7 planar foils (could be more if necessary). The specially designed stacks of radiochromic films (RCFs) and BAS-SR image plates (IPs), located 6.3 cm behind the rear of the target, are used to measure the profile and energy spectrum of protons and electrons, respectively. These stacks are of 50 mm×50 mm large, with a 3-mm-diameter hole in the center to let protons pass through, and enwrapped by 15 μm-thickness Al foil to shield debris. Copper and aluminum sheets of different thickness are inserted between RCFs and SR-IPs to attenuate proton and electron energy, respectively. Due to limited space only two stacks are used simutaneously in an experiment.

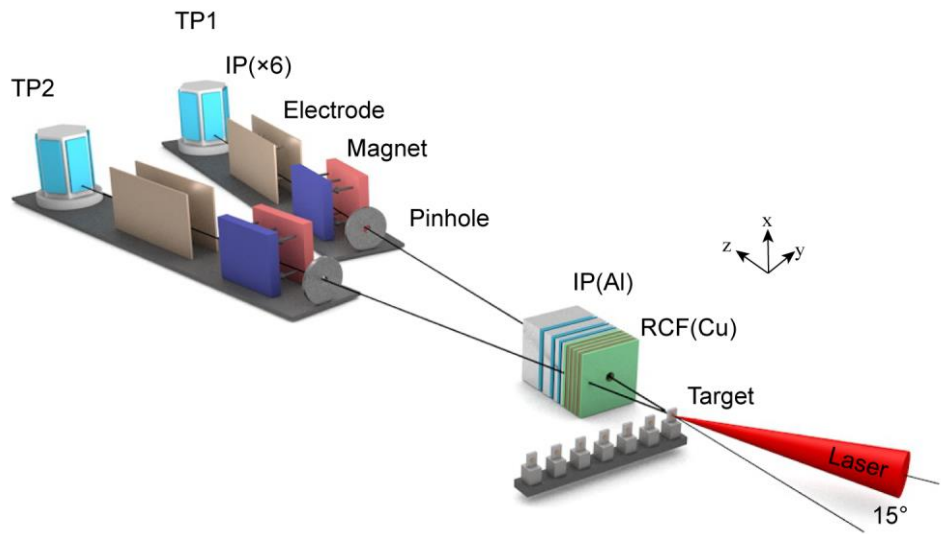

Fig. 4. The sketch of the experimental setup. The specially designed stacks of radiochromic films and BAS-SR image plates are used to measure the profile and energy spectrum of protons and electrons. The stacks and targets can move along *y* direction. Two Thomson parabola spectrometers are used to detect the ion spectra at the target normal direction and laser direction. It installs six BAS-TR image plates at a time.

Two types of Thomson parabola (TP) spectrometers (TP1 and TP2) are used to detect proton energy spectrum, as shown in Fig. 4. The TP1 composed of 1.0 T magnetic field over 5-cm long and a pair of 15-cm long copper electrodes charged up to 10 kV is placed 87.8 cm away from the target along the target normal direction. The diameter of TP1's pinhole is 150 μm corresponding to the solid angle of $2.3 \times 10^{-8}$ sr. The energy resolution of TP1 is 0.4 MeV @ 100 MeV, with a low energy threshold of 3.5 MeV. The other high-resolution TP2, placed 80 cm away from the target along the laser direction, is switched on when RCFs and IPs are not in use. The TP2 employs magnetic field of 1.7 T over 5-cm long and electrodes up to 35 cm long with 10 kV. It has a pinhole of 200 μm diameter, corresponding to $4.9 \times 10^{-8}$ sr solid angle. The energy resolution reaches 0.13 MeV @ 100 MeV and the lower energy threshold is 9.2 MeV for TP2. From Fig. 4 one should note that the ions cannot be detected by TP2 if the stacks move in. In both TPs, BAS-TR IPs are placed at a turntable holder that can rotate 360 degrees in the horizontal plane, allowing for six successive measurements without interruption.

The whole TPs are surrounded by a lead shield to reduce the signal noise during experiment. For future development, an online detector named microchannel plate (MCP) with a fluorescent screen [30] will be installed to improve the diagnosis efficiency.

### 3.2 Acceleration of 60 MeV proton beam vis TNSA

The maximum energy of protons accelerated by intense ultrafast lasers is mainly determined by, but not limited to, the laser intensity, pulse duration and pulse contrast ratio, and thus is considered as an important perspective to find out a laser facility's capabilities. The most widely studied mechanisms for laser-driven ion acceleration are TNSA [31,32] and radiation pressure acceleration (RPA) [33,34], which require different laser and target parameters. In the commissioning experiment, under the current conditions of SULF-10 PW described above, we focus on TNSA using the micrometer-thick Cu foils.

The proton cut-off energy as a function of target thickness is shown in Fig. 5(a), measured by TP1 along the target normal direction and by both TP2 and RCF stacks along the laser propagation direction. The target thickness $l$ of the Cu-foil varies from 1 μm to 10 μm. It can be clearly seen that in both directions, the proton cut-off energy increases when the foil thickness increases from 1 μm to 4 μm, and then decreases with larger thickness, corresponding to an optimum value at 4 μm. This trend agrees with the previously reported results for TNSA-produced proton beams, where the effects of electron reflux and pre-pulse induced plasma expansion at the target rear side results in an optimum target thickness for proton acceleration [35]. One notices that the proton energies along the target normal direction are much higher than those along the laser direction for all target thicknesses. Typical proton spectra for various target thicknesses are illustrated in Fig. 5(b), showing the broad-energy-spread distribution.

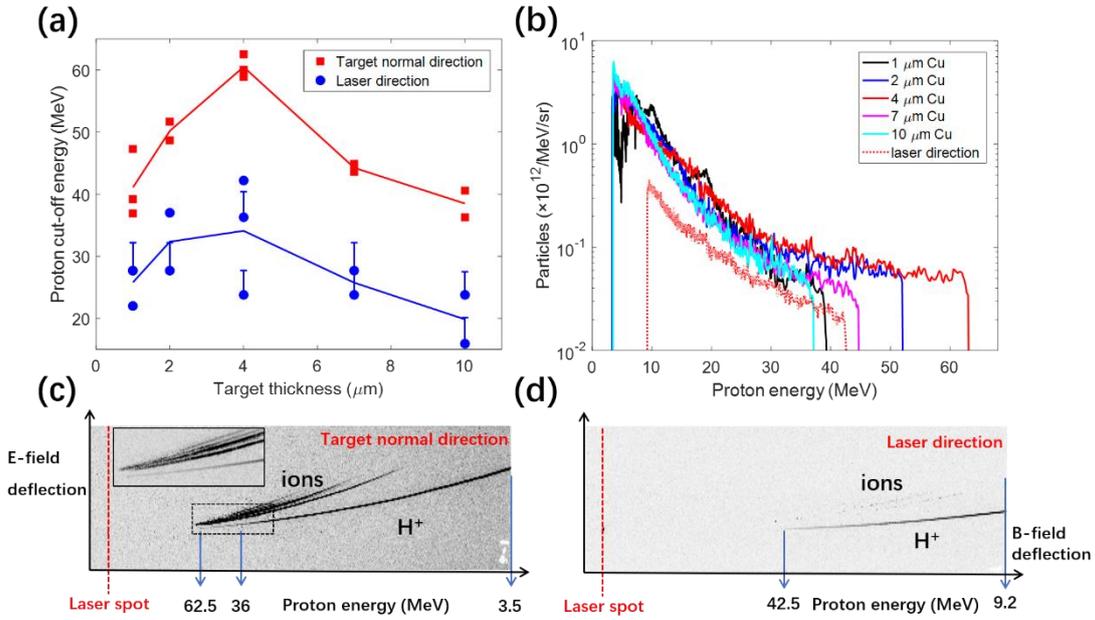

Fig. 5. (a) The proton cut-off energy as a function of the target thickness of the plain Cu foils measured by TP1 in the target normal direction (red squares) and by both TP2 and RCF stacks in the laser propagation direction (blue circles), where the red and blue lines represent the average proton energy over 2-3 shots. The vertical error bars for some data are defined by the energy interval between adjacent RCF layers. (b) Typical proton spectra for five target thicknesses of $l$ = 1 μm (black line), 2 μm (blue line), 4 μm (red line), 7 μm (magenta line) and 10 μm (cyan line) in the target normal direction, respectively. The proton energy spectrum for $l$ = 4 μm (dashed red line) in the laser direction is also included in (b). (c)-(d) The raw IP data of TP1 and TP2 for the best result of proton acceleration from a shot on a 4-μm Cu foil, where the inset in (c) is a magnified image of the ion trace in the high-energy region.

For 4-μm-thick foils, the average cut-off energy of protons is 60 MeV according to the data from 3 shots, where the highest one achieves 62.5 MeV. This is among the state-of-art results in proton acceleration using femotosecend lasers according to the previous reports [31,36]. Figures 5(c)-(d) show the raw IP data of TP1 and TP2 for the best case with the 4-μm Cu target, from which the proton energy spectra are extracted and presented in Fig. 5(b). Note that both the cut-off energy and particle number of protons in the target normal direction are much higher than those measured along the laser propagation direction. The cut-off energies are 42.5 MeV (laser direction) and 62.5 MeV (target normal direction), respectively. From the proton spectrum of the best shot (see in Fig. 5(b)), the proton number for energy more than 3.5 MeV can be estimated and reaches up to $2.4 \times 10^{12}$, corresponding to 1.5% energy conversion efficiency (assuming the proton beam with divergence half-angle of 10°).

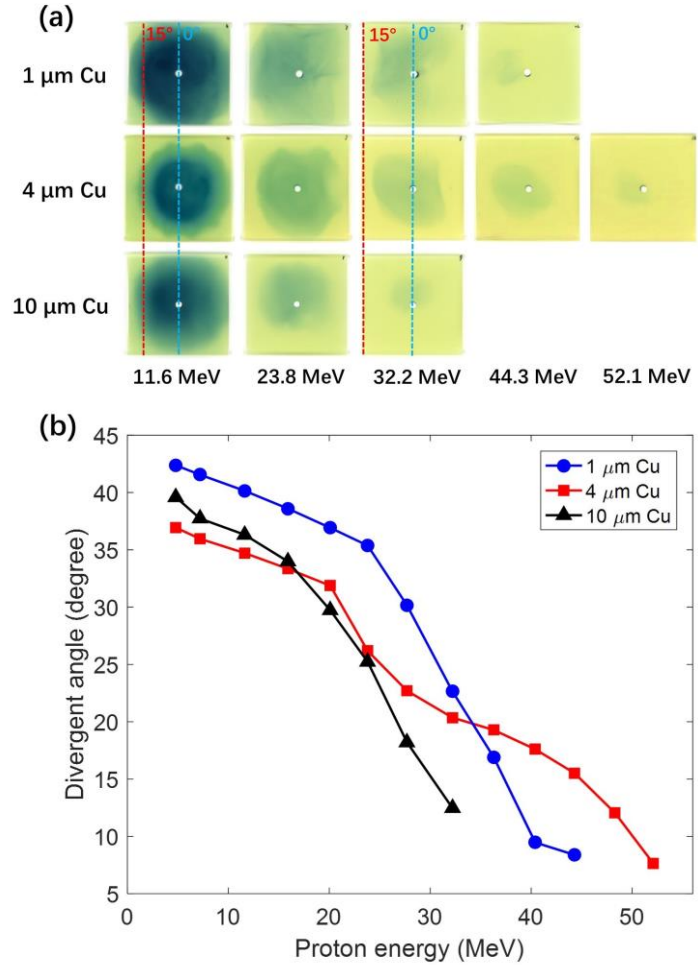

Fig.6. (a) Typical proton profiles from three shots on Cu target of $l$ = 1 μm, 4 μm and 10 μm, at selected layers of RCF stacks corresponding to the proton energies of 11.6 MeV, 23.8 MeV, 32.2 MeV, 44.3 MeV and 52.1 MeV, respectively. The target normal direction (0°) and laser direction (15°) are illustrated by dashed blue and red lines for 11.6 MeV and 32.2 MeV. (b) Divergent angle of protons at different energies for $l$ = 1 μm (blue circles), 4 μm (red squares), and 10 μm (black triangles).

Typical proton profiles from three shots on Cu target of $l$ = 1 μm, 4 μm and 10 μm are shown in Fig. 6(a) at selected layers of RCF stacks, with the highest energy detected of 44.3 MeV, 52.1 MeV and 32.2 MeV, respectively. These results are slightly lower than the actual values according to signal of TP1 with the cut-off energies of 47.3 MeV, 58.9 MeV and 36.3 MeV for the same shots. This is mainly due to not only the large interval of energy measurement between adjacent RCF layers but also the use of the HD-V2 type of RCFs which cannot detect the low-density protons in the high-energy region. It can be found from Fig. 6(a) that the proton signal monotonically decreases at higher energies. Figure 6(b) shows the corresponding divergent angle of proton beam as a function of the energy. Protons are more collimated at higher energies. The minimum divergences measured for $l$ = 1 μm, 4 μm, and 10 μm are 8.4°, 7.6° and 12.5°, respectively. It is a typical feature in the TNSA regime, which is different from that of RPA-produced protons where the beam is more divergent at higher proton energies [37,38]. From the density profiles of protons >32.2 MeV, it can be found that the center of proton beams is not exactly aligned with the target normal direction, but shifts slightly toward the laser direction in the case of $l$ = 1 μm and 4 μm (see in Fig. 6(a)). This is mainly due to the bending of target surface induced by the laser pre-pulse before the main pulse arrives [39].

The electron number distributions are also measured using IP stacks that record electrons of energy over 11.8 MeV, 14.2 MeV, 17.2 MeV, 20.2 MeV and 23.7 MeV (electrons with energies lower than 11 MeV are stopped by the RCF stacks), respectively. As shown in Fig. 7(a), the emitting direction of electrons also shift slightly toward the laser direction (>0°), similar to the proton beam (see in Fig. 6(a)). Considering the total electron signal within each IP shown in Fig. 7(a), the electron numbers within four energy intervals of 11.8-14.2 MeV, 14.2 -17.2 MeV, 17.2-20.2 MeV and 20.2-23.7 MeV are obtained and the processed spectrum is displayed in Fig. 7(b). The fitting curve of electron spectrum indicates an electron temperature $T_e$ of 7.6 MeV, which agrees reasonably well with the theoretical result of 8.6 MeV following the ponderomotive scaling law given by $T_e(m_e c^2) = \int_0^{\frac{2\pi}{w_0}} \sqrt{1 + a_0^2 \sin^2(w_0 t)}\, dt/(2\pi/w_0) - 1$ [40], where $m_e$, $c$, $a_0$ and $w_0$ are the electron mass, light speed, peak normalized vector potential and angular frequency of the laser pulse, respectively (the laser intensity for this shot is $1.7 \times 10^{21}$ W/cm$^2$). The features mentioned above suggest that TNSA is dominant given the provided laser intensity and micrometer-thick Cu foils. Both the 60 MeV proton beams of several shots and the electron temperature reflect the current capabilities of the SULF-10 PW beamline for providing laser beams with ultrahigh intensity, ultrashort duration and relatively high contrast.

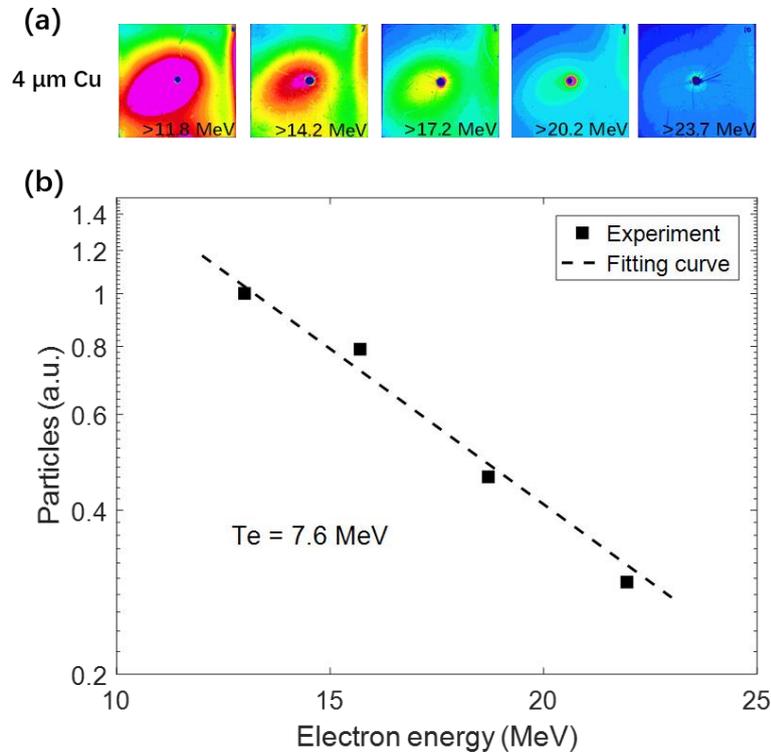

Fig.7. (a) Electron number distribution measured using IP stacks for electron energy greater than 11.8 MeV, 14.2 MeV, 17.2 MeV, 20.2 MeV and 23.7 MeV, from the same shot on a 4-μm-thick Cu target as illustrated in Fig. 6. (b) The processed electron spectrum, where the dashed line represents the fitting curve.

### 3.3 Proton acceleration using nanometer-thick targets

Laser-driven proton acceleration using nanometer-thick plastic (CH) foils is also investigated here. The profiles of proton beams at selected energies for CH foils with target thicknesses of $l$=30 nm, 40 nm and 70 nm are exhibited in Fig. 8. The highest proton energies shown by the RCF are smaller than those measured by TP1 with the cut-off energies of 19.2 MeV, 20.0 MeV and 19.4 MeV for $l$=30 nm, 40 nm and 70 nm, respectively. Clear ring-like profiles appear for $l$=30 nm and 40 nm on all RCF layers, which is probably induced by relativistic transparency effect in nanometer target situation [37]. In both cases, the divegent angles of protons remain almost unchanged at different energies and the center of proton beams is well aligned along the target normal direction, as shown in Figs. 8(a)-(b), in contrast to the TNSA case using micrometer-thick targets [39] and the RPA case [37]. Such properties indicate that protons are not effectively accelerated since ionizaion and pre-expanding of nanometer-thick targets driven by pre-pulses may lead to relativistically transparent plasma [41].

Filamented structure emerges when the target thickness increases to 70 nm (see in Figs. 8(c1)-(c4)), which is possibly associated with Weibel instability [42] or the wrinkles on the target surface [43]. The divegent angles of protons become smaller for more energetic protons while the center of the proton beam profile mainly concentrates near the laser propagation axis. It is an obvious sign that the plasma is still opaque rather than transparent. Considering the use of linearly polarized lasers and an oblique incidence angle of 15°, the proton acceleration for $l$=70 nm case may be dominant by a hybrid scheme where both the hole boring stage [44] of RPA and TNSA play important roles. The results with nanometer-thick targets show that the laser contrast of SULF-10 PW beamline is not sufficient to drive effective acceleration schemes such as RPA [33,34] and acceleration using structural target [45,46]. Proton energies beyond 100 MeV are expected after further optimization of the temporal contrast and focal spot of SULF-10 PW beamline.

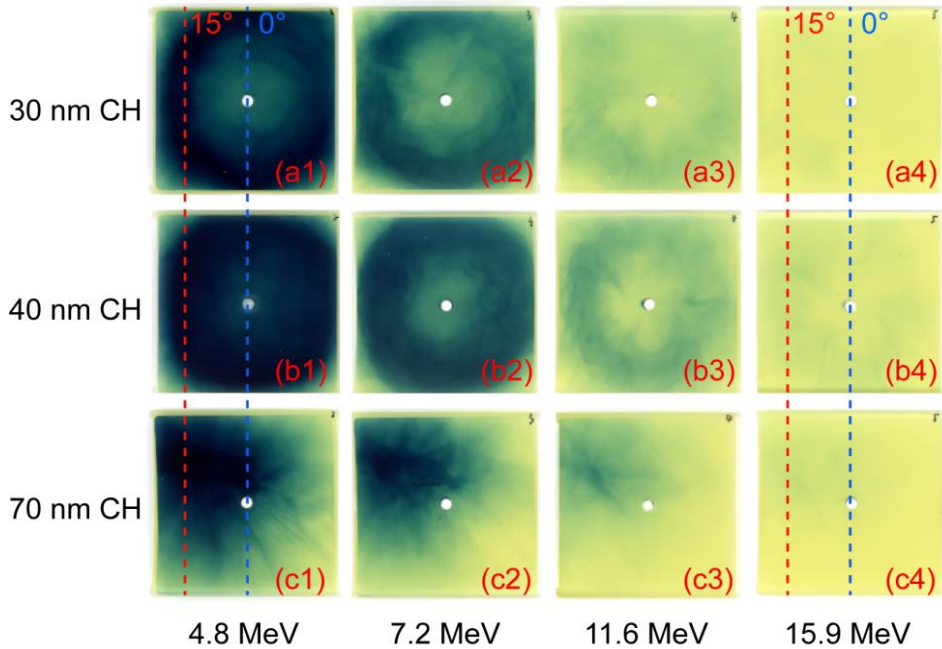

Fig.8. Proton beam profiles for plain CH targets with three different thicknesses of (a1-a4) 30 nm, (b1-b4) 40 nm and (c1-c4) 70 nm, at selected proton energies of 4.8 MeV, 7.2 MeV, 11.6 MeV and 15.9 MeV, respectively. The dashed lines in blue and red represent the target normal direction (0°) and laser direction (15°), respectively.

## 4. Conclusions and perspectives

A commissioning experiment of SULF-10 PW beamline has been carried out, focusing on laser-proton acceleration with the plain Cu and plastic targets. The SULF-10 PW laser beamline can provide 2.4 PW peak power on target currently. A high-energy proton beam with maximum cut-off energy up to 62.5 MeV was achieved with Cu foils at the optimum target thickness of 4 μm via TNSA, which is approaching to the requirement of tumor therapy treatment [47]. For plastic targets of tens-of-nanometer thickness, the proton profiles show apparent ring-like or filamented structures. The experimental results illustrate the current status of the SULF-10 PW beamline.

The on-target peak power of the SULF-10 PW beamline will be increased to 10 PW after maintenance of the pump sources in the last amplifier. Further optimization works to improve laser intensity and contrast are continuing through using the smaller $f$- number OAP and setting up a traditional plasma mirror. In the near future, the peak laser intensity is expected to reach $10^{22}$-$10^{23}$ W/cm$^2$, which provides strong support for research in strong field physics.

## Acknowledgments

This work was supported by the Strategic Priority Research Program of the Chinese Academy of Sciences (Grant No. XDB16), the National Natural Science Foundation of China (Grant Nos. 11875307, 11935008, 11804348, 11705260, 11905278 and 11975302), and Youth Innovation Promotion Association of Chinese Academy of Science (No. 2021242).